\documentclass[prd, reprint, floatfix, nolongbibliography, superscriptaddress]{revtex4-2}

\usepackage{hyperref} 
\usepackage{amsmath}
\usepackage{amsthm}
\usepackage{amssymb}
\usepackage{mathtools}
\usepackage[english]{babel}

\usepackage{graphicx}
\usepackage{float}
\usepackage{pgf}

\renewcommand{\vec}{\boldsymbol}
\newcommand{\ds}{\displaystyle}

\definecolor{amendments}{rgb}{0.0, 0.0, 0.7}

\begin{document}

\thispagestyle{empty}

\title{Total yield of electron-positron pairs produced from vacuum in strong electromagnetic fields: validity of the locally constant field approximation}
\author{D.~G.~Sevostyanov}
\affiliation{Department of Physics, St.~Petersburg State University, 7/9 Universitetskaya Naberezhnaya, Saint Petersburg 199034, Russia}
\author{I.~A.~Aleksandrov}
\email{i.aleksandrov@spbu.ru}
\affiliation{Department of Physics, St.~Petersburg State University, 7/9 Universitetskaya Naberezhnaya, Saint Petersburg 199034, Russia}
\affiliation{Ioffe Institute, Politekhnicheskaya 26, Saint Petersburg 194021, Russia}
\author{G.~Plunien}
\affiliation{Institut f\"ur Theoretische Physik, Technische Universit\"at Dresden, Mommsenstrasse 13, Dresden D-01062, Germany \vspace{10mm}}
\author{V.~M.~Shabaev}
\affiliation{Department of Physics, St.~Petersburg State University, 7/9 Universitetskaya Naberezhnaya, Saint Petersburg 199034, Russia}

\begin{abstract}
\vspace{-1cm}
The widely-used locally constant field approximation (LCFA) can be utilized in order to derive a simple closed-form expression for the total number of particles produced in the presence of a strong electromagnetic field of a general spatio-temporal configuration. A usual justification for this approximate approach is the requirement that the external field vary slowly in space and time. In this investigation, we examine the validity of the LCFA by comparing its predictions to the results obtained by means of exact nonperturbative numerical techniques. To benchmark the LCFA in the regime of small field amplitudes and low frequencies, we employ a semiclassical approach. As a reference, we consider a standing electromagnetic wave oscillating both in time and space as well as two spatially uniform field configurations: Sauter pulse and oscillating electric field. Performing a thorough numerical analysis, we identify the domain of the field parameters where the approximation is well justified. In particular, it is demonstrated that the Keldysh parameter is not a relevant quantity governing the accuracy of the LCFA.
\end{abstract}

\maketitle

\section{Introduction}\label{sec:intro}

One of the most remarkable phenomena predicted by quantum electrodynamics (QED) is the process of electron-positron pair production in strong external fields~\cite{sauter_1931, euler_heisenberg, Weisskopf, schwinger_1951}. Even in the absence of real particles, vacuum fluctuations of the quantum electron-positron field may be galvanized due to the interaction with a strong external background, so the vacuum state decays producing $e^+ e^-$ pairs. In a constant uniform electric field $E_0$, the probability of this striking phenomenon is proportional to $\mathrm{exp} (-\pi E_\text{c}/E_0)$, where $E_\text{c} = m^2 c^3/(|e|\hbar) \approx 1.3\times 10^{16}~\text{V/cm}$ is a critical (Schwinger) value of the electric field strength. This fundamental process cannot be described by means of perturbation theory. As the amplitudes of the most intense laser pulses that can be generated nowadays are far below $E_\text{c}$, one may expect that the Schwinger pair production will not be experimentally observed within the near future. On the other hand, in nonstationary backgrounds the temporal oscillations of the field can effectively enhance the probability of the process, even if its nonperturbative nature is still preserved. Furthermore, taking into account the pre-exponential factor could also boost the particle yield within the theoretical estimations. To obtain reliable quantitative predictions, one basically strives to consider realistic field configurations. However, due to the limited computational resources, it is unfeasible to perform nonperturbative calculations for general inhomogeneous backgrounds in (3+1) dimensions. To reduce the computational time when studying complex scenarios, one can employ approximate methods instead of carrying out rigorous calculations.

A widely-used approach is based on the idea that the spatio-temporal scale of quantum processes is commonly much smaller than the characteristic parameters of the external background. It allows one to treat the field as {\it locally} constant and utilize simple results derived in the case of space-time-independent backgrounds. The actual implementation of this general approach should be discussed once the physical process is specified. For instance, in the context of nonlinear Compton scattering (NCS)~\cite{nikishov_ritus_1964, brown_kibble_1964, narozhny_nikishov_ritus_1965}, i.e., emission of a photon by an electron or positron in a strong field, and the so-called nonlinear Breit-Wheeler (NBW) mechanism~\cite{breit_wheeler, reiss_1962, narozhny_nikishov_ritus_1965}, i.e., decay of a photon into an electron-positron pair, the external field can be considered locally constant if the ``formation length'' of the quantum processes is much smaller than the length scale of the electromagnetic background, e.g., laser wavelength. Moreover, one can employ the closed-form expressions for the corresponding rates derived in the case of constant crossed fields~\cite{nikishov_ritus_1964, ritus_1985}. This approach provides a powerful tool for studying strong-field QED phenomena in arbitrary space-time-dependent backgrounds. Note that the applicability of the locally constant field approximation (LCFA) depends not only on the external field parameters but also on the initial state of the electron (for NCS) or photon (for NBW). The validity of the LCFA in the context of NCS and NBW is discussed in recent papers~\cite{di_piazza_lcfa_2018, di_piazza_lcfa_2019, ilderton_lcfa_2019, king_ulcfa_2020, seipt_king_2020} (see also references therein).

In the present study, we will examine the LCFA being applied to the process of vacuum pair production. If one aims to approximately calculate the momentum distributions of particles created, a consistent formulation of the LCFA exists in the case of spatially uniform external fields~\cite{aleksandrov_prd_2019_1}, whereas for taking into account both temporal and spatial inhomogeneities, there is no rigorous LCFA procedures, so one can only propose ad hoc approximate methods such as, e.g., the local dipole approximation~\cite{aleksandrov_prd_2019_1, aleksandrov_kohlfuerst} treating locally only the spatial dependence of the field. In contrast, the {\it total} number of pairs can be evaluated within the LCFA by using the exact expression for the pair-creation rate in a constant field~\cite{nikishov_jetp_1970,bagrov_1975}, where space and time are considered in the same way. Since this expression is valid for constant electric and magnetic fields of arbitrary direction and magnitude, one can simply integrate it over space and time taking into account the inhomogeneities of the external background under consideration (see, e.g., Refs.~\cite{bunkin_tugov, narozhny_pla, narozhny_jetp, bulanov_prl_2010, gavrilov_prd_2017}). As a constant electromagnetic field produces pairs in the tunneling regime, one can expect the LCFA to be well justified only in the domain of slowly varying fields with small Keldysh parameters $\gamma = (E_\text{c}/E_0)[\hbar \omega/(mc^2)]$, where $\omega$ is the frequency of the external field. The condition $\gamma \ll 1$ is equivalent to the requirement that the formation length $2mc^2/(|e|E_0)$ be much smaller than the laser wavelength $\lambda = 2\pi c/\omega$. The main goal of the present study is to investigate the accuracy of the LCFA once $\gamma \neq 0$.

To this end, one has to compare the LCFA predictions with the exact results obtained by means of rigorous QED techniques. Note that we are interested in the process of vacuum pair production in a classical background treated nonperturbatively within the zeroth order in the fine-structure constant, i.e., we neglect the radiative corrections, so ``exact methods'' are those dealing with the spatio-temporal inhomogeneities of the external field without any additional approximations. First, we will employ the so-called quantum kinetic equations (QKE)~\cite{GMM, schmidt_1998, kluger_1998} to analyze spatially uniform scenarios. To take into account the coordinate dependence in the case of a standing electromagnetic wave, we will use our numerical technique~\cite{aleksandrov_prd_2016, aleksandrov_prd_2018, aleksandrov_kohlfuerst} based on the Furry-picture quantization of the electron-positron field~\cite{fradkin_gitman_shvartsman}. Integrating the momentum distributions of electrons created, we compute the total number of pairs in order to benchmark the LCFA and deduce general patterns concerning its justification. Unfortunately, numerical methods can be effectively applied only within the domain of sufficiently large field amplitudes ($E_0 \gtrsim 0.1 E_\text{c}$) and high frequencies ($\hbar \omega \gtrsim 0.1 mc^2$). To assess the accuracy of the LCFA also in the case of smaller $E_0$ and $\omega$, we will perform a semiclassical analysis based on the imaginary-time method~\cite{ppt2, popov_jetp_lett_1971, popov_jetp_1971, popov_jetp_1972, popov_yad_fiz_1974}. This allows us to gain a complete picture of where the LCFA is valid.

The paper is organized as follows. In Sec.~\ref{sec:LCFA} we discuss the implementation of the LCFA for computing the total number of pairs and provide the explicit formulas for two specific scenarios: oscillating electric field and standing electromagnetic wave. In Sec.~\ref{sec:exact} we outline the exact techniques utilized in the study. Section~\ref{sec:results} contains the main findings in the case of spatially homogeneous field configurations. In Sec.~\ref{sec:semiclass} we perform a semiclassical analysis of an oscillating electric field. In Sec.~\ref{sec:sw} we examine a standing electromagnetic wave varying both in time and space. Finally, we conclude in Sec.~\ref{sec:conclusion}.

Throughout the article, we assume $\hbar = c = 1$ and the electron charge $e < 0$.

\section{Locally constant field approximation}\label{sec:LCFA}

The idea of the LCFA is to employ the exact closed-form expression for the total number of pairs created in a constant uniform electromagnetic field. Let us first recall the main results concerning pair production in this simple external background.

Due to the interaction with the external classical electromagnetic field, the initial vacuum state of the quantized electron-positron field may not remain a vacuum state after the interaction if the amplitude of the vacuum-vacuum transition has an absolute value less than unity. Let us denote this amplitude by $\mathrm{e}^{iW}$. If the effective action $W$ has a nonzero imaginary part, then the vacuum decay probability is also nonzero and reads
\begin{equation}\label{eq:decay_action_im}
  \mathcal{P}_\text{decay} = 1 - | \mathrm{e}^{iW} |^2 = 
  1 - \mathrm{e}^{-2 \mathrm{Im} \, W}.
\end{equation}
In the case of a constant external field, the effective action is a product of the spatio-temporal volume $VT$ and the effective Lagrangian $\mathcal{L}$. The latter is the one-loop effective Lagrangian $\mathcal{L}^{(\text{1})}$~\cite{euler_heisenberg, Weisskopf, schwinger_1951} since we disregard the quantized part of the electromagnetic field. The corresponding imaginary part~\cite{schwinger_1951} is given by
\begin{equation}\label{eq:eff_L_im}
  2 \, \mathrm{Im} \, \mathcal{L}^{(\text{1})} [\mathcal{E}, \mathcal{H} ] = \frac{e^2 \mathcal{E} \mathcal{H}}{4\pi^2} \sum_{n=1}^\infty \frac{1}{n} \coth{\frac{n \pi \mathcal{H}}{\mathcal{E}}} \, \mathrm{exp} \bigg (\! - \frac{n \pi m^2}{|e|\mathcal{E}} \bigg ),
\end{equation}
where $\mathcal{E}$ and $\mathcal{H}$ are defined via
\begin{eqnarray}
  \mathcal{E} &=& \sqrt{\sqrt{\mathcal{F}^2 + \mathcal{G}^2} + \mathcal{F}},\label{eq:E3_cal} \\
  \mathcal{H} &=& \sqrt{\sqrt{\mathcal{F}^2 + \mathcal{G}^2} - \mathcal{F}}.\label{eq:H_cal}
\end{eqnarray}
Here $\mathcal{F} = (\vec{E}^2 - \vec{H}^2)/2$, $\mathcal{G} = \vec{E}\cdot \vec{H}$, and the vectors $\vec{E}$ and $\vec{H}$ are the electric and magnetic field in a given (laboratory) frame. The invariant quantities $\mathcal{E}$ and $\mathcal{H}$ coincide with the magnitudes of the fields $\vec{E}$ and $\vec{H}$, respectively, if these vectors are parallel to each other. According to Eq.~\eqref{eq:eff_L_im}, the imaginary part is negligible unless the electric field strength $\mathcal{E}$ is close to the critical value $E_\text{c} = m^2/|e|$.

The vacuum instability appears due to the process of electron-positron pair production. In the present study, we are interested in calculating the total number of pairs $N$, which in the case of a constant background can be found exactly~\cite{nikishov_jetp_1970} (see also Ref.~\cite{bagrov_1975}). If the external field does not vary in space and time, it is the number of pairs {\it per unit volume and time} which yields a finite value, $N/(VT)$. One can also view it as a differential quantity $dN/(dtd\boldsymbol{x})$, and its explicit form reads~\cite{nikishov_jetp_1970,bagrov_1975}
\begin{equation}
  \frac{dN}{dt d\vec{x}} \, [\mathcal{E}, \mathcal{H} ] = \frac{e^2 \mathcal{E} \mathcal{H}}{4\pi^2} \coth{\frac{\pi \mathcal{H}}{\mathcal{E}}} \, \mathrm{exp} \bigg ( \! - \frac{\pi m^2}{|e|\mathcal{E}} \bigg ).
  \label{eq:N_density}
\end{equation}
Note that this expression is the first term of the series~\eqref{eq:eff_L_im} and its physical meaning differs from that of $2 \, \mathrm{Im} \, \mathcal{L}$ (see, e.g., Refs.~\cite{nikishov_fian_1979, fradkin_gitman_shvartsman, cohen_2008}). Nevertheless, if the external field produces a small number of pairs according to Eq.~\eqref{eq:N_density}, then Eqs.~\eqref{eq:decay_action_im}, \eqref{eq:eff_L_im}, and \eqref{eq:N_density} yield indistinguishable values. As we are interested in computing the total number of pairs, in what follows we will employ Eq.~\eqref{eq:N_density}.

If the external field slowly varies in space and time, one can approximately treat it as locally constant, i.e., assume that the particle number arising from a small space-time region of size $\Delta V \Delta T$ around the point $x = (t, \vec{x})$ can be evaluated by means of Eq.~\eqref{eq:N_density} where $\mathcal{E}$ and $\mathcal{H}$ are replaced with $\mathcal{E} (x)$ and $\mathcal{H} (x)$, respectively. Multiplying the density by $\Delta V \Delta T$ and summing such individual contributions over space and time, one immediately receives a Riemann sum which for $\Delta V$, $\Delta T \to 0$ yields
\begin{equation}
  N^{(\text{LCFA})} = \int \! d^4x \, \frac{dN}{dt d\vec{x}} \, [\mathcal{E} (x), \mathcal{H} (x) ].
  \label{eq:N_LCFA_gen}
\end{equation}
The LCFA treats time and the spatial coordinates in a uniform manner. If the external field exists in a finite space-time domain, the integral in Eq.~\eqref{eq:N_LCFA_gen} provides a finite dimensionless value, which is a total number of pairs produced.

Our aim is to clarify what the expression ``slowly varying field'' actually means, i.e., to find out when the LCFA formula~\eqref{eq:N_LCFA_gen} is justified. In what follows, we will discuss how this approximate method is used in the case of two specific field configurations.

\subsection{Standing electromagnetic wave}

First, we will discuss how the LCFA approach can be implemented in the case of a standing wave. The external field is described in the gauge $A_0 = 0$ by the following vector potential:
\begin{equation}\label{eq:field_config}
  \vec{A} (t, z) = \frac{E_0}{\omega}\, F(t) \sin \omega t \cos k_z z \, \vec{e}_x, 
\end{equation}
where $k_z = \omega$, $\vec{e}_x$ is a unit vector along the $x$ direction, and $F(t)$ is a smooth envelope function vanishing for $t \to \pm \infty$. The electric and magnetic fields then read
\begin{eqnarray}
  \vec{E} (t, z) &=& -E_0 \left [ F(t) \cos \omega t + \frac{F'(t)}{\omega} \, \sin \omega t \right ] 
  \cos k_z z \, \vec{e}_x, \quad\;\,\, \label{eq:field_E} \\ 
  \vec{H} (t, z) &=& -E_0 F(t) \sin \omega t \, \sin k_z z \, \vec{e}_y.\label{eq:field_H}
\end{eqnarray}
Plugging them into the definition of $\mathcal{F}$ and $\mathcal{G}$, we obtain $\mathcal{G} = 0$ and
\begin{widetext}
  \begin{equation} \label{eq:field_F}
    \mathcal{F} (t, z) = \frac{E_0^2}{2} F^2 (t) \cos(\omega t + k_z z) \cos(\omega t - k_z z) 
    + \frac{E_0^2}{2} \, \frac{F'(t)}{\omega} 
    \left[ F(t) \sin 2 \omega t + \frac{F'(t)}{\omega} \, \sin^2 \omega t \right] \cos^2 k_z z.
  \end{equation}
\end{widetext}
We choose the envelope in the following form:
\begin{equation}
  F (t)=
  \begin{cases}
    \sin^2  \ds \frac{\pi N_\text{c} - |\omega t|}{2} &\text{if}~~\pi (N_\text{c}-1) \leqslant |\omega t| < \pi N_\text{c},\\
    1 &\text{if}~~|\omega t| < \pi (N_\text{c}-1),\\
    0 &\text{otherwise},
    \end{cases} \label{eq:envelope_explicit}
\end{equation}
so it contains a flat plateau of $N_\text{c} - 1$ carrier cycles and two switching on/off parts of half a cycle each. In what follows, we will use $N_\text{c} = 5$.

Let us now present the explicit form of Eq.~(\ref{eq:N_LCFA_gen}) in the case of the standing wave~(\ref{eq:field_config})--(\ref{eq:field_H}). First, since $\mathcal{G} = 0$, one of the two invariants $\mathcal{E}$ and $\mathcal{H}$ is always zero: 
\begin{eqnarray}
\mathcal{E} = \sqrt{2\mathcal{F}},~\mathcal{H} = 0 &~~~& \text{for}~~~\mathcal{F} \geqslant 0, \label{eq:F_pos} \\
\mathcal{E} = 0,~\mathcal{H} = \sqrt{-2\mathcal{F}} &~~~& \text{for}~~~\mathcal{F} < 0. \label{eq:F_neg}
\end{eqnarray}
The space-time regions where $\mathcal{F}$ is negative do not contribute to $N^{(\text{LCFA})}$. For $\mathcal{F} \geqslant 0$, we take the limit $\mathcal{H} \to 0$ in Eq.~\eqref{eq:N_density}. Second, integration over the $xy$ plane leads to a factor $S$ representing the cross section of the system, which is assumed to tend to infinity. Third, as the external field is also infinite in the $z$ direction and periodic with a period of $2\pi/k_z = 2\pi/\omega$, integration over $z$ should be performed according to the prescription
\begin{equation}
\int \! dz = \int \limits_{-L_z/2}^{L_z/2} \!\!\!\!\! dz = \int \limits_{0}^{2\pi/\omega} \!\!\! dz \, \frac{\omega L_z}{2\pi},
\label{eq:integral_dz}
\end{equation}
where $L_z$ is the size of the system in the $z$ direction, so $V = S L_z$. From this, it follows that the total number of pairs can be evaluated as
\begin{widetext}
  \begin{equation}\label{eq:N_LCFA_SW}
    N^{(\text{LCFA})} = \frac{\omega V}{8\pi^4} \int \! dt \!\! \int \limits_{0}^{2\pi/\omega} \!\! dz \, \theta (\vec{E}^2 - \vec{H}^2) \, e^2 (\vec{E}^2 - \vec{H}^2) \, 
    \mathrm{exp} \left( \! -\frac{\pi m^2}{|e|\sqrt{\vec{E}^2 - \vec{H}^2}} \right) ,
  \end{equation} 
\end{widetext}
where $\theta (x)$ is the Heaviside step function and the fields $\vec{E}$ and $\vec{H}$ are given by Eqs.~\eqref{eq:field_E} and \eqref{eq:field_H}, respectively. In the final estimations, the full volume of the system $V$ can be replaced with the interaction volume depending on how the external laser pulses are focused. In order to evaluate the integrals in Eq.~(\ref{eq:N_LCFA_SW}), it is convenient to use the substitutions $\phi = \omega t$ and $\psi = \omega z$. It immediately becomes clear that the product $\omega N^{(\text{LCFA})}$ is completely independent of $\omega$, provided the envelope function depends only on $\omega t$ (e.g., the pulse duration is governed by the number of cycles which is independent of $\omega$). Thus we will factor out the $\omega$ dependence as well as the volume $V$ presenting the results in terms of the following quantity:
\begin{equation}
  \nu^{\text{(LCFA)}} = \frac{\omega}{V} N^{\text{(LCFA)}}.
\label{eq:nu_def}
\end{equation}
To benchmark the LCFA results, we will also perform exact calculations of the analogous function $\nu = \omega N/V$ depending on $\omega$ beyond the LCFA.

\subsection{Spatially homogeneous oscillating electric field}\label{sec:oef_LCFA}

We will also consider a purely time-dependent field of the following form:
\begin{equation}
  \vec{A}(t) = \frac{E_0}{\omega} F(t) \sin \omega t \, \vec{e}_x.
  \label{eq:field_config_oef}
\end{equation}
The field invariant $\mathcal{F}$ can be obtained by setting $z=0$ in Eq.~\eqref{eq:field_F} and it is given by
\begin{eqnarray}
    \mathcal{F} (t) &=& \frac{E_0^2}{2} F^2 (t) \cos^2 \omega t \nonumber \\ 
    {}&+& \frac{E_0^2}{2} \, \frac{F'(t)}{\omega} \left [ F(t)\sin 2 \omega t + \frac{F'(t)}{\omega} \, \sin^2 \omega t \right ].
\end{eqnarray}
In Eq.~\eqref{eq:N_LCFA_gen} we replace $\mathcal{E}$ with $E(t) = -A'_x (t)$ and again take the limit $\mathcal{H} \to 0$ in Eq.~\eqref{eq:N_density}. The number of pairs per unit volume is then determined via
\begin{equation}
  \frac{N^{\text{(LCFA)}}}{V} = \frac{e^2}{4\pi^3} \int \! dt \,  E^2(t) \, \mathrm{exp} 
  \left( \! - \frac{\pi m^2}{|eE(t)|} \right).
  \label{eq:LCFA_uniform_final}
\end{equation}
This expression being multiplied by $\omega$ is also independent of $\omega$, so we present the results in terms of the function $\nu^{\text{(LCFA)}}$ defined by Eq.~\eqref{eq:nu_def}.

\section{Exact approaches}\label{sec:exact}

To explore the validity of the LCFA, we benchmark this approximation against the exact methods. In this section, we briefly describe two different approaches. The first one is based on the so-called kinetic description of vacuum pair production in the case of spatially homogeneous fields. The second method rests on the Furry-picture quantization of the electron-positron field in the presence of a space-time-dependent external background. If the field depends solely on time, the two methods yield identical results, whereas the latter approach allows us to analyze the pair-creation process in a standing electromagnetic wave taking into account its spatial inhomogeneity.

\subsection{Quantum kinetic equations}

In the case of a spatially homogeneous external field, a nonperturbative kinetic approach was developed in Refs.~\cite{GMM, schmidt_1998, kluger_1998} and subsequently generalized in Refs.~\cite{pervushin_2006, filatov_2006} allowing one to consider uniform fields of arbitrary polarization (see also recent review~\cite{aleksandrov_epj_2020} and references therein). Here we will recap the basics of the approach in the case of a linearly polarized electric field.

We assume that the external field is directed along the $x$ axis:
\begin{equation}
  \vec{A}(t) = A(t) \vec{e}_x, \quad \vec{E}(t) = E(t) \vec{e}_x.
\end{equation}
The field vanishes outside the interval $t \in [t_\text{in}, \, t_\text{out}]$. For instance, Eq.~\eqref{eq:envelope_explicit} implies $t_\text{out} = -t_\text{in} = \pi N_\text{c}/\omega$.

The process of pair production is described in terms of the electron number density in momentum space,
\begin{equation}
  f(\vec{p}) = \frac{(2\pi)^3}{V} \frac{dN_{\vec{p},s}}{d\vec{p}}.
  \label{eq:f_dist_def}
\end{equation}
It does not carry the spin quantum number $s$ because in the linearly-polarized field the particle distribution does not depend on $s$. To obtain the total number of pairs, one should integrate $f(\vec{p})$ over $\vec{p}$ and multiply the result by $2$ due to the spin degeneracy. Within the quantum kinetic approach, one introduces a time-dependent distribution function $f(\vec{p},t)$ which obeys $f(\vec{p},t_\text{out}) = f(\vec{p})$. At the initial time instant $t = t_\text{in}$, we employ the vacuum condition $f(\vec{p},t_\text{in}) = 0$. The evolution of the distribution function is governed by the following integro-differential equation:
\begin{equation}\label{eq:vlasov}
  \dot{f}(\vec{p},t) = \lambda (\vec{p},t) \!
  \int \limits_{t_\text{in}}^t \! dt' \, \lambda (\vec{p}, t')
  \left[\frac{1}{2} - f(\vec{p},t') \right]
  \cos 2\theta(t,t'),
\end{equation}
where
\begin{eqnarray}
  && \lambda (\vec{p},t) = \frac{eE(t)\mu}{\omega^2(\vec{p}, t)}, \\
  && \mu = \sqrt{m^2+p_{\perp}^2}, \\
  && \omega(\vec{p}, t) = \sqrt{\mu^2 + [p_{\parallel}-eA(t)]^2}, \\
  && \theta(t,t') = \int \limits^{t}_{t'} \!\omega(\vec{p}, t'') \, d t''. 
\end{eqnarray}
Due to the azimuthal symmetry, the distribution function depends only on the longitudinal momentum projection $p_\parallel = p_x$ and the transverse one $p_\perp = \sqrt{p_y^2 + p_z^2}$. Alternatively, one can recast Eq.~\eqref{eq:vlasov} into the following Cauchy problem:
\begin{eqnarray}\label{eq:cauchy_1}
\dot{f}(\vec{p},t) &=& \frac{1}{2}\lambda u(\vec{p},t),\\
\dot{u}(\vec{p},t) &=& \bigl[1-2f(\vec{p},t)\bigr] - 2\omega v(\vec{p},t), \label{eq:cauchy_2} \\
\dot{v}(\vec{p},t) &=& 2\omega u(\vec{p},t). \label{eq:cauchy_3}
\end{eqnarray}
Here $\lambda$ and $\omega$ depend on $\vec{p}$ and $t$, and we assume $f(\vec{p}, t_{\text{in}}) = u(\vec{p}, t_{\text{in}}) = v(\vec{p}, t_{\text{in}}) = 0$ for all $\vec{p}$. The system~\eqref{eq:cauchy_1}--\eqref{eq:cauchy_3} is usually referred to as the quantum kinetic equations (QKE).

The QKE can be solved numerically for given $\vec{p}$, i.e., $p_\parallel$ and $p_\perp$. Varying $\vec{p}$, one obtains the momentum spectra of the electrons created. In this study, we integrate the distribution function over $\vec{p}$ ($d\vec{p} \to 2 \pi p_\perp d p_\perp dp_\parallel$) to compute the total number of pairs.

\subsection{Furry picture in momentum space}\label{sec:furry}

To incorporate both the temporal and spatial inhomogeneities of the external field, one should employ more general methods than that discussed in the previous subsection. For instance, the so-called Dirac-Heisenberg-Wigner formalism is a kinetic approach allowing one, in principle, to investigate arbitrary backgrounds in (3+1) dimensions~\cite{vasak_ann_phys_1987, BB_1991}. However, we will use a general technique based on the Furry-picture quantization with the aid of in/out solutions~\cite{fradkin_gitman_shvartsman} and discuss its implementation within momentum space~\cite{aleksandrov_prd_2016}.

Let us consider the Dirac equation including the interaction with a classical background $A^\mu = (A_0, \vec{A})$:
\begin{equation} \label{eq:dirac_general}
  \big ( \gamma^\mu \big [i \partial_\mu - e A_\mu (t, \vec{x}) \big ] - m \big) \psi (t, \vec{x}) = 0.
\end{equation}
We define two sets of solutions ${}_\zeta \psi_n (t, \vec{x})$ and ${}^\zeta \psi_n (t, \vec{x})$ which satisfy the following conditions:
\begin{equation}\label{eq:psi_in_out}
  {}_\zeta \psi_n (t_\text{in}, \vec{x}) = {}_\zeta \psi^{(0)}_n (\vec{x}),\quad {}^\zeta \psi_n (t_\text{out}, \vec{x}) = {}^\zeta \psi^{(0)}_n (\vec{x}).
\end{equation}
Here ${}_\zeta \psi^{(0)}_n (\vec{x})$ and ${}^\zeta \psi^{(0)}_n (\vec{x})$ form two complete sets of the Hamiltonian eigenfunctions at $t = t_\text{in}$ and $t = t_\text{out}$, respectively, with the sign of the energy denoted by $\zeta = \pm$. The functions ${}_\zeta \psi_n (t, \vec{x})$ and ${}^\zeta \psi_n (t, \vec{x})$ are called {\it in}- and {\it out}-solutions, respectively. The number density of electrons created with quantum numbers $l$ then reads~\cite{fradkin_gitman_shvartsman}
\begin{equation}
  n^-_l = \sum_n  G({}^+|{}_-)_{ln} G({}_-|{}^+)_{nl}, \label{eq:num_el}
\end{equation}
where the $G$ matrices are defined as the following inner products:
\begin{eqnarray}
  G({}^\zeta|{}_\varkappa)_{ln} &=& ({}^\zeta \psi_l, \, {}_\varkappa \psi_n), \label{eq:G_inner_product_1}\\
  G({}_\zeta|{}^\varkappa)_{ln} &=& ({}_\zeta \psi_l, \, {}^\varkappa \psi_n). \label{eq:G_inner_product_2}
\end{eqnarray}
Note that they do not depend on time because the Dirac Hamiltonian is Hermitian. According to Eq.~\eqref{eq:num_el}, the external field creates pairs only if it permits transitions between the positive-energy and negative-energy continua.

The asymptotic behavior of the in- and out-solutions is given by the simple plane-wave solutions of the Dirac equation for a free particle. These states are determined by momentum $\vec{p}$ and spin quantum number $s = \pm 1$, i.e., $l=\{\vec{p}, s\}$. Instead of evaluating the inner products~\eqref{eq:G_inner_product_1} and \eqref{eq:G_inner_product_2} in coordinate space, one can evolve the Fourier components of the electron wavefunction which directly yield the elements of the corresponding $G$ matrices. This approach is described in detail in Ref.~\cite{aleksandrov_prd_2016}.

The method is particularly efficient in the case of spatially periodic backgrounds~\cite{aleksandrov_prd_2016, aleksandrov_prd_2018, aleksandrov_kohlfuerst}. If the external field is described by the vector potential $\vec{A}(t,\vec{x}) = A_x(t,z) \, \vec{e}_x,$ and $A_x(t,z)$ is a periodic function of $z$ with period $2\pi/\omega$, then the out-solution can be represented in the following form:
\begin{equation}\label{eq:fourier_def}
  {}^+ \psi_{\vec{p}, s} (t, \vec{x}) = \frac{\mathrm{e}^{i\vec{p}\vec{x}}}{(2\pi)^{3/2}} \, \chi_{\vec{p},s}(t,z),
\end{equation}
where
\begin{equation}
\chi_{\vec{p},s}(t,z) = 
  \sum_{j = -\infty}^{+\infty} \mathrm{e}^{i\omega j z} \, w^{j}_{\vec{p},s}(t).
\end{equation}
In terms of the four-component time-dependent functions $w^{j}_{\vec{p},s} (t)$, the Dirac equation reads
\begin{equation}
  i\dot{w}^{j}_{\vec{p},s} = ( \vec{\alpha} \cdot \vec{p} + \alpha_z \omega j + \beta m ) w^{j}_{\vec{p},s} - 
  e\alpha_x \!\! \sum_{l=-\infty}^{\infty} a_{l-j}(t)w^{l
  }_{\vec{p},s}, 
\end{equation}
where $a_l$ are the Fourier components of the external field,
\begin{equation}
  a_l(t) = \frac{\omega}{2\pi}\int_{0}^{2\pi/\omega} A_x(t,z) \mathrm{e}^{-i\omega l z} dz.
\end{equation}
Taking into account the ``initial'' conditions $w^j_{\vec{p},s}(t_\text{out}) = u_{\vec{p}, s} \delta_{j0}$, where $u_{\vec{p}, s}$ is a constant bispinor corresponding to the positive-energy states, we evolve the functions $w^j_{\vec{p},s}(t)$ backwards in time and obtain the density~\eqref{eq:f_dist_def} via
\begin{equation}
f(\vec{p}) = \sum_{j=-\infty}^{\infty} \sum_{s'=\pm 1} 
  \big|v^{\dag}_{p_x, p_y, p_z+\omega j, \, s'} w^{j}_{\vec{p},s}(t_{\text{in}}) \big|^2.
\label{eq:f_furry_final}
\end{equation}
Here we have used Eqs.~\eqref{eq:num_el}--\eqref{eq:G_inner_product_2}; $v_{\vec{p}, s}$ is a constant bispinor regarding the negative-energy wavefunction. Equation~\eqref{eq:f_furry_final} yields the same result no matter what value of $s$ is chosen. A detailed comparison between this numerical approach and that based on the Dirac-Heisenberg-Wigner formalism can be found in Ref.~\cite{aleksandrov_kohlfuerst}.

\section{Results: Spatially homogeneous external fields}\label{sec:results}

In this section, we will benchmark the LCFA against the exact results obtained with the aid of the QKE and Furry-picture formalism for two different scenarios involving spatially homogeneous backgrounds: Sauter pulse and oscillating electric field.
\subsection{Sauter pulse}
The external field is purely electric and reads
\begin{equation}
  E_x(t) = \ds \frac{E_0}{\cosh^2 (t/\tau)}, \quad E_y = E_z = 0.
 \label{eq:sauter_field}
\end{equation}
A crucial feature of this field configuration is that the number density of electrons produced can be derived analytically~\cite{narozhny_1970}:
\begin{widetext}
  \begin{equation}
    \frac{(2\pi)^3}{V} \frac{d N_{\vec{p},s}}{d\vec{p}}
    = \frac{\sinh \Big [ \frac{1}{2} \pi \tau (2e E_0 \tau + \omega_- - \omega_+)\Big ] \sinh 
    \Big [ \frac{1}{2} \pi \tau (2e E_0 \tau + \omega_+ - \omega_-)\Big ]}{\sinh(\pi \omega_+ \tau) 
    \sinh(\pi \omega_- \tau)},\label{eq:LCFA_sauter_exact}
  \end{equation}
\begin{figure*}[b]
    \center{\includegraphics[width=0.95\textwidth]{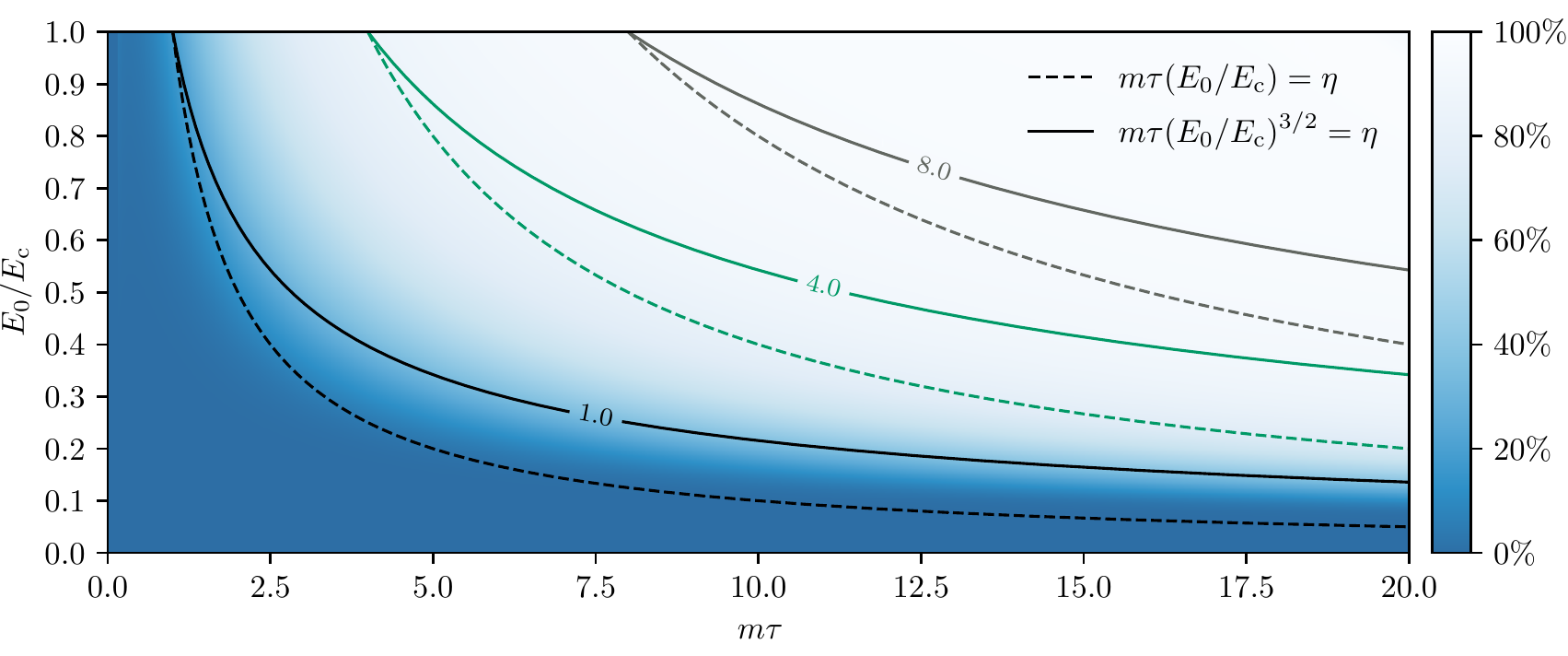}}
    \caption{Ratio $[N^{\text{(LCFA)}}/N] \times 100\%$ as a function of the amplitude $E_0$ of a Sauter pulse~\eqref{eq:sauter_field} and its duration $\tau$. The three color-coded pairs of curves correspond to $\eta = 1.0$, $4.0$, and $8.0$, where $\eta$ is defined in the legend separately for solid and dashed lines.}
    \label{fig:sauter}
\end{figure*}  
\end{widetext}
where $\omega_{\pm} = \ds \sqrt{\mu^2 + (P_\parallel \mp eE_0\tau)^2}$, $\mu^2 = m^2 + p_{\bot}^2$, and 
$P_\parallel = p_\parallel - eE_0 \tau$. Numerically integrating this expression, we evaluated the total particle yield and tested our numerical procedure developed for treating arbitrary time-dependent external fields.

The LCFA is accurate only if the Sauter pulse has a sufficiently large duration $\tau$ and amplitude $E_0$. For instance, the requirement that the Keldysh parameter be much smaller than unity, $\gamma \ll 1$, is equivalent to the following condition:
\begin{equation}
  |eE_0|\tau \gg m.
  \label{eq:LCFA_sauter_cond_keldysh}
\end{equation}
We assume that the characteristic frequency of the Sauter pulse is $\omega = 1/\tau$. Nevertheless, as was demonstrated in Ref.~\cite{aleksandrov_prd_2019_1}, the relevant condition of the LCFA applicability has a different form:
\begin{equation}
  |eE_0|^{3/2}\tau \gg m^2.
  \label{eq:LCFA_sauter_cond}
\end{equation}
This result was obtained with the assumption $m\tau \gg 1$, which is completely realistic since in relativistic units $m^{-1} = 1.3 \times 10^{-21}~\text{s}$. To demonstrate which of the inequalities~\eqref{eq:LCFA_sauter_cond_keldysh} and \eqref{eq:LCFA_sauter_cond} is relevant to the LCFA justification, we directly compare the approximate results for the number of pairs with the exact values.

As the LCFA incorporates only the tunneling mechanism completely neglecting multiphoton processes, it always underestimates the particle yield, i.e., $N^{\text{(LCFA)}} < N$, so it is convenient to present the results in terms of the quantity $[N^{\text{(LCFA)}}/N] \times 100\%$ (see Fig.~\ref{fig:sauter}). We also depict the three pairs of curves determined by $\eta = m \tau (E_0/E_\text{c})$ (dashed lines) and $\eta = m \tau (E_0/E_\text{c})^{3/2}$ (solid lines), where $\eta = 1.0$, $4.0$, and $8.0$ for the black, green, and gray curves, respectively. The heatmap confirms that the relevant condition is given by Eq.~\eqref{eq:LCFA_sauter_cond} as, unlike the dashed line, the solid line ``1.0'' is very close to a contour line of the graph (as was indicated above, we are not interested in small values of $m\tau$). The naive condition~\eqref{eq:LCFA_sauter_cond_keldysh} is weaker than~\eqref{eq:LCFA_sauter_cond}, so the LCFA may not be valid even if the Keldysh parameter is small.
\subsection{Oscillating electric field}\label{sec:oef_num}
\begin{figure}[t]
  \center{\includegraphics[width=0.95\linewidth]{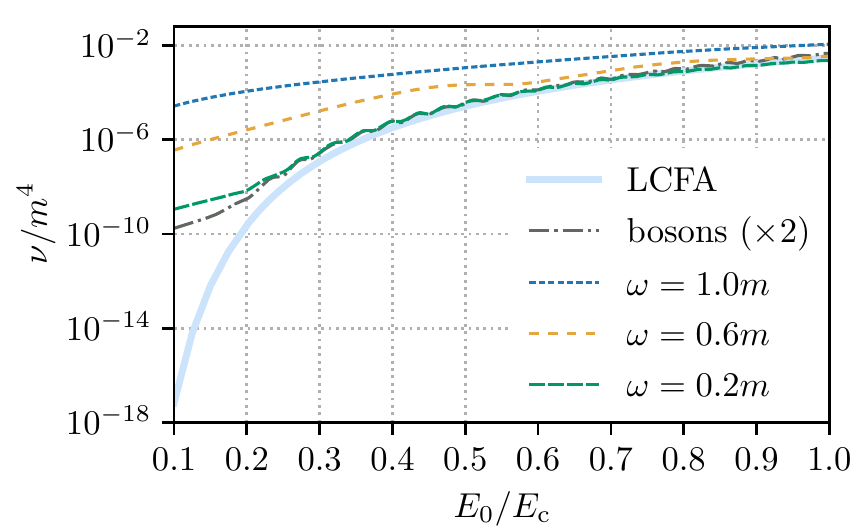}}
  \caption{Parameter $\nu = \omega N/V$ as a function of the amplitude of the oscillating background~\eqref{eq:field_config_oef} calculated within the LCFA (solid light-blue line) and computed exactly for various values of the field frequency $\omega$ (dashed lines). The dash-dotted line shows the exact results obtained in the case of scalar particles (the data was multiplied by 2 in order to balance the Fermi spin factor).}
  \label{fig:oef}
\end{figure}

Here we will discuss the LCFA in the case of a spatially homogeneous oscillating electric field~\eqref{eq:field_config_oef}. As was pointed out above, the results will be presented in terms of the function $\nu = \omega N/V$ generally depending on $E_0$ and $\omega$. The LCFA yields, however, $\omega$-independent results. We compare them with the exact predictions in Fig.~\ref{fig:oef}. The LCFA accurately reproduces the results of complete quantum simulations if the external field is sufficiently strong and its frequency is low. Evaluating the Keldysh parameter, we observe that the green dashed line ($\omega = 0.2m$) is far from the LCFA solid line at $E_0 = 0.2 E_\text{c}$, i.e., $\gamma = 1.0$, whereas the results almost coincide at $E_0 = 0.4 E_\text{c}$, i.e., $\gamma = 0.5$. Moreover, other values of $\omega$ lead to different threshold values of $\gamma$, so the condition $\gamma \ll 1$ seems again irrelevant to the problem of the LCFA justification.

\begin{figure}[t]
  \center{\includegraphics[width=0.95\linewidth]{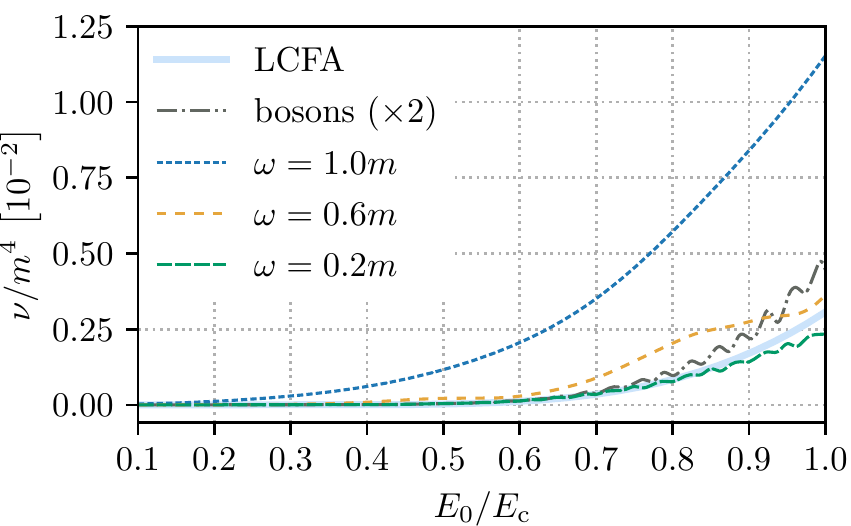}}
  \caption{Parameter $\nu = \omega N/V$ versus the amplitude of the oscillating background~\eqref{eq:field_config_oef} evaluated within the LCFA (solid light-blue line) and computed exactly for various $\omega$ (dashed lines). The dash-dotted line shows the exact results obtained in the case of scalar particles.}
  \label{fig:oef_zoom}
\end{figure}

As was mentioned above, the LCFA is likely to underestimate the particle yield. Nevertheless, the green line in Fig.~\ref{fig:oef} ($\omega = 0.2m$) intersects the LCFA curve. To make it evident, we display the same graph with a linear scale (see Fig.~\ref{fig:oef_zoom}). We observe that the LCFA results become greater than the exact values for $E_0 \gtrsim 0.85 E_\text{c}$. The physical explanation for this behavior lies in the fact that the LCFA does not take into account the Pauli exclusion principle at all, while the distribution function $f(\vec{p})$ computed exactly cannot exceed unity. For sufficiently large field amplitudes, many of the positive-energy electron states become occupied, which inhibits the pair-production process. To confirm this statement, we performed the analogous calculations in the case of bosons, i.e., within scalar QED (see the gray dash-dotted line in Figs.~\ref{fig:oef} and \ref{fig:oef_zoom}). First, note that the LCFA in the case of bosons leads to precisely the same expressions~\eqref{eq:N_density} and \eqref{eq:N_LCFA_gen}, provided the boson yield is multiplied by the Fermi spin factor 2 (see, e.g., Ref.~\cite{fradkin_gitman_shvartsman}), although the imaginary part of the effective Lagrangian is different~\cite{Weisskopf, schwinger_1951, fradkin_gitman_shvartsman}. Second, the QKE method should be modified according to the following prescription~\cite{GMM, kluger_1998}: one has to change the sign of $f(\vec{p}, t')$ in the square brackets in Eq.~\eqref{eq:vlasov} and define $\lambda = eE(t)[p_{\parallel} - eA(t)]/\omega^2$. According to our data, the LCFA results never exceed the total number of bosons as shown in Figs.~\ref{fig:oef} and \ref{fig:oef_zoom}.

Finally, we underline that the exact values of the total number of pairs decreases much more slowly with decreasing $E_0$ compared to the LCFA predictions. Moreover, one should also keep in mind that the interaction volume in the real experimental setup may yield a huge additional factor since the Compton volume $m^{-3}$ corresponds to $5.76 \times 10^{-38}~\text{m}^3$. For instance, one cubic micrometer, which seems a realistic interaction volume, is $19$ orders of magnitude larger. This suggests that the actual threshold of pair production can be considerably lower than $E_\text{c}$ (see, e.g., Refs.~\cite{narozhny_pla, narozhny_jetp}).

In the regime of weaker fields and lower frequencies, where $E_0 \ll E_\text{c}$ and $\omega \ll m$, we are not able to carry out the exact computations due to the technical limitations. On the other hand, one can invoke semiclassical methods to benchmark the LCFA, which will be discussed in the next section.

\section{Semiclassical analysis} \label{sec:semiclass}

Semiclassical methods are expected to be accurate once $E_0 \ll E_\text{c}$ and $\omega \ll m$. In the case of a spatially homogeneous oscillating background $\boldsymbol{E} (t) = E_0 \sin \omega t \, \boldsymbol{e}_x$, the problem was analyzed, e.g., in Refs.~\cite{popov_jetp_lett_1971, popov_yad_fiz_1974} by means of the so-called imaginary time method~\cite{ppt2, popov_jetp_1971, popov_jetp_1972}. In what follows, we will first consider the limit $\omega \to 0$ for given $E_0$ and compare the LCFA results with the semiclassical predictions. Second, we will turn to the case of nonzero $\omega$ to determine to which extent one can rely on the LCFA if pair production is no longer a pure tunneling process.
\subsection{Zero-frequency limit}
Here in Sec.~\ref{sec:semiclass}, we do not introduce the envelope function as it is inessential for the present analysis. In this case, evaluating the LCFA expression~\eqref{eq:LCFA_uniform_final}, we obtain the following number of pairs per unit spatio-temporal volume:
\begin{equation}
 \frac{N_{\infty}^{\text{(LCFA)}}}{VT} = \frac{m^4}{2\pi^4} \, \bigg ( \frac{E_0}{E_\text{c}} \bigg )^2 \int \limits_0^{\pi/2} \! d \phi \, \cos^2 \phi \ \mathrm{exp} \left( \! - \frac{\pi E_\text{c}}{E_0 \cos \phi} \right),
 \label{eq:LCFA_uniform_wo_envelope}
\end{equation}
where the subscript ``$\infty$'' indicates that the external pulse is infinite in time, i.e., there is no envelope. Note that the expression~\eqref{eq:LCFA_uniform_wo_envelope} does not depend on $\omega$ at all, so it can be considered as the limit of the {\it exact result} as $\omega \to 0$ for given $E_0$ as long as the Pauli principle is insignificant, i.e., the field strength $E_0$ is not too large.

It is convenient to use the Keldysh parameter,
\begin{equation}
\gamma = \frac{m \omega}{|eE_0|} = \frac{E_\text{c}}{E_0} \, \frac{\omega}{m}.
 \label{eq:Keldysh}
\end{equation}
The zero-$\gamma$ limit of the exact results is given by Eq.~\eqref{eq:LCFA_uniform_wo_envelope}. Let us compare it with the semiclassical estimates. In the limit $\gamma \to 0$, the imaginary-time method (ITM) yields~\cite{popov_jetp_lett_1971, popov_yad_fiz_1974} (see also Ref.~\cite{ringwald_2001})
\begin{equation}
 \frac{N_{\infty}^{\text{(SC)}}}{VT} = \frac{m^4}{2^{3/2} \pi^4} \, \bigg ( \frac{E_0}{E_\text{c}} \bigg )^{5/2} \mathrm{exp} \left( \! - \frac{\pi E_\text{c}}{E_0} \right).
 \label{eq:LCFA_uniform_wo_envelope_SC_zero_gamma}
\end{equation}
Since we are interested in the total number of pairs, we can exploit the ITM expression, although it does not take into account the interference effects that are pronounced in the momentum spectra of particles produced in laser pulses with a subcycle structure~\cite{dumlu_prl_2010, dumlu_prd_2011_1}. The result~\eqref{eq:LCFA_uniform_wo_envelope_SC_zero_gamma} should coincide with Eq.~\eqref{eq:LCFA_uniform_wo_envelope} once $E_0 \ll E_\text{c}$. To explicitly verify this, we first express the integral in Eq.~\eqref{eq:LCFA_uniform_wo_envelope} as $\mathcal{J} (\pi E_\text{c}/E_0)$, where
\begin{figure}[t]
  \centering
  \includegraphics[width=\linewidth]{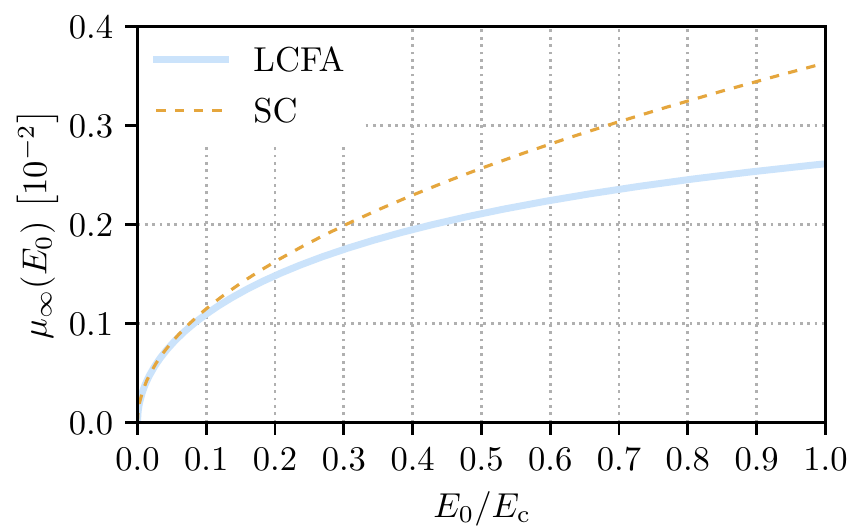}
  \caption{Function $\mu_\infty (E_0)$ defined in Eq.~\eqref{eq:LCFA_uniform_mu_def} and computed within the LCFA [Eq.~\eqref{eq:mu_lcfa}, solid line] and by means of the semiclassical expression~\eqref{eq:mu_SC} (dashed line).}
  \label{fig:mu_E0}
\end{figure}
\begin{equation}
\mathcal{J} (z) = \frac{z}{2} \, \int \limits_1^{\infty} \bigg ( 1 + \frac{1}{x^2} \bigg ) K_0 (zx) dx.
 \label{eq:LCFA_J_def}
\end{equation}
This representation is derived in the Appendix. Assuming then $z \gg 1$, i.e., $E_0 \ll E_\text{c}$, one can deduce the asymptotic expansion of the LCFA result. It is convenient to use the dimensionless function $\mu_\infty (E_0)$ defined by
\begin{equation}
 \frac{N_{\infty}}{VT} = m^4 \bigg ( \frac{E_0}{E_\text{c}} \bigg )^{2} \mathrm{exp} \left( \! - \frac{\pi E_\text{c}}{E_0} \right) \mu_\infty (E_0).
 \label{eq:LCFA_uniform_mu_def}
\end{equation}
Then we receive
\begin{eqnarray}
\mu_\infty^{\text{(SC)}} (E_0) &=& \frac{1}{2^{3/2} \pi^4} \, \sqrt{\frac{E_0}{E_\text{c}}}, \label{eq:mu_SC} \\
\mu_\infty^{\text{(LCFA)}} (E_0) &=& \frac{1}{2\pi^4} \, \mathrm{exp} \left( \frac{\pi E_\text{c}}{E_0} \right) \mathcal{J} \bigg ( \frac{\pi E_\text{c}}{E_0} \bigg ).
\label{eq:mu_lcfa}
\end{eqnarray}
The asymptotic behavior of the LCFA expression~\eqref{eq:mu_lcfa} for $E_0 \ll E_\text{c}$ reads (see the Appendix)
\begin{equation}
\frac{\mu_\infty^{\text{(LCFA)}} (E_0)}{\mu_\infty^{\text{(SC)}} (E_0)} =  1 - \frac{13}{8\pi} \frac{E_0}{E_\text{c}} + \frac{657}{128\pi^2} \frac{E_0^2}{E_\text{c}^2} + \mathcal{O} \bigg ( \frac{E_0^3}{E_\text{c}^3} \bigg ) ,
 \label{eq:LCFA_mu_asym}
\end{equation}
so for $E_0 \ll E_\text{c}$ we recover the semiclassical prediction. The constant term in Eq.~\eqref{eq:LCFA_mu_asym} can also be identified by expanding $1/\cos \phi \approx 1 + \phi^2/2$ in the exponential in Eq.~\eqref{eq:LCFA_uniform_wo_envelope} and integrating then the Gaussian function over $\phi \in (0, \, \infty)$. Note that the derivation presented in the seminal paper of Brezin and Itzykson~\cite{BI_1970} yields $\mu_\infty = 1/(8\pi^2)$, so it does not properly capture the pre-exponential factor.

To find out how strict the condition $E_0 \ll E_\text{c}$ is, we plot Eqs.~\eqref{eq:mu_SC} and \eqref{eq:mu_lcfa} versus $E_0$ (see Fig.~\ref{fig:mu_E0}). Note that the total particle yield is proportional directly to $\mu_\infty (E_0)$ according to Eq.~\eqref{eq:LCFA_uniform_mu_def}, so these two quantities have the same relative uncertainties. In Fig.~\ref{fig:mu_E0} we observe that the semiclassical approach is quite accurate even if $E_0$ is greater than $0.1 E_\text{c}$. If $E_0 = 0.1 E_\text{c}$, the relative error of the semiclassical result amounts to $4.9 \%$. Note that taking into account the terms up to $(E_0/E_\text{c})^2$ in Eq.~\eqref{eq:LCFA_mu_asym}, one obtains the LCFA result for $E_0 = 0.1E_\text{c}$ with a relative uncertainty of $0.06\%$.

Having examined the validity of the semiclassical approach in the limit $\gamma \to 0$, i.e. $\omega \to 0$, we now turn to the analysis of the LCFA justification for nonzero $\gamma$.
\subsection{Nonzero field frequency}
The semiclassical approach also allows one to obtain the total number of pairs produced for nonzero values of the Keldysh parameter $\gamma$. For instance, if $\gamma \gg 1$, instead of Schwinger's exponential $\mathrm{exp} \, (-\pi E_\text{c} / E_0)$, one receives a power-law multiphoton behavior of the particle yield, which can be deduced by means of perturbation theory. Since the LCFA takes into account only the tunneling mechanism of pair production, one can expect this approximation to be accurate only within a certain vicinity of $\gamma = 0$ [recall that the LCFA prediction~\eqref{eq:LCFA_uniform_wo_envelope} is $\gamma$-independent for given $E_0$]. In order to determine the size of this vicinity, we expand the general semiclassical expressions given in Ref.~\cite{popov_yad_fiz_1974} in terms of $\gamma$. Instead of Eq.~\eqref{eq:mu_SC}, we obtain
\begin{eqnarray}
\mu_\infty^{\text{(SC)}} (E_0, \gamma) &=& \mu_\infty^{\text{(SC)}} (E_0) \, \Big [ 1 + \mathcal{O}(\gamma^2) \Big ] \nonumber \\
{}& \times & \mathrm{exp} \left[ \! \frac{\pi E_\text{c}}{E_0} \, \frac{\gamma^2}{8} \bigg \{ 1 + \mathcal{O} (\gamma^2)  \bigg \}  \right].
 \label{eq:mu_SC_gen}
\end{eqnarray}
In the limiting case, $\mu_\infty^{\text{(SC)}} (E_0, 0)  = \mu_\infty^{\text{(SC)}} (E_0)$. We assume that $E_0 < 0.1 E_\text{c}$, so $\mu_\infty^{\text{(LCFA)}} (E_0) \approx \mu_\infty^{\text{(SC)}} (E_0)$ with a relative uncertainty of less than $5\%$. To make sure that the field frequency is sufficiently low, so $\mu_\infty^{\text{(SC)}} (E_0, \gamma)  \approx \mu_\infty^{\text{(SC)}} (E_0)$ and the LCFA is applicable, we should require
\begin{equation}
\frac{E_\text{c}}{E_0} \, \gamma^2 \ll 1,
 \label{eq:LCFA_SC_cond_1}
\end{equation}
which is equivalent to
\begin{equation}
\frac{|eE_0|^{3/2}}{\omega} \gg m^2.
 \label{eq:LCFA_SC_cond_2}
\end{equation}
This condition is exactly the same as Eq.~\eqref{eq:LCFA_sauter_cond} for $\tau = 1/\omega$. Note that the inequality~\eqref{eq:LCFA_SC_cond_2} is, in fact, quite weak due to extracting the square root of Eq.~\eqref{eq:LCFA_SC_cond_1} as well as the factor $\pi/8$ in the exponential in Eq.~\eqref{eq:mu_SC_gen}. For instance, if we require $\mu_\infty^{\text{(SC)}} (E_0, \gamma) < 1.1 \, \mu_\infty^{\text{(SC)}} (E_0)$, i.e., the relative uncertainty does not exceed $10\%$, then the field amplitude and its frequency should obey $|eE_0|^{3/2} > 2.0 \, \omega m^2$.

To summarize, if $E_0 < 0.1E_\text{c}$, then the LCFA is valid, provided Eq.~\eqref{eq:LCFA_SC_cond_2} is satisfied. This condition is considerably stronger than the naive requirement $\gamma \ll 1$ once $E_0 \lesssim 0.1E_\text{c}$. If $E_0 > 0.1 E_\text{c}$, the LCFA predictions can be compared with the exact numerical results as was discussed in the previous section.

\begin{figure}[t]
  \center{\includegraphics[width=0.95\linewidth]{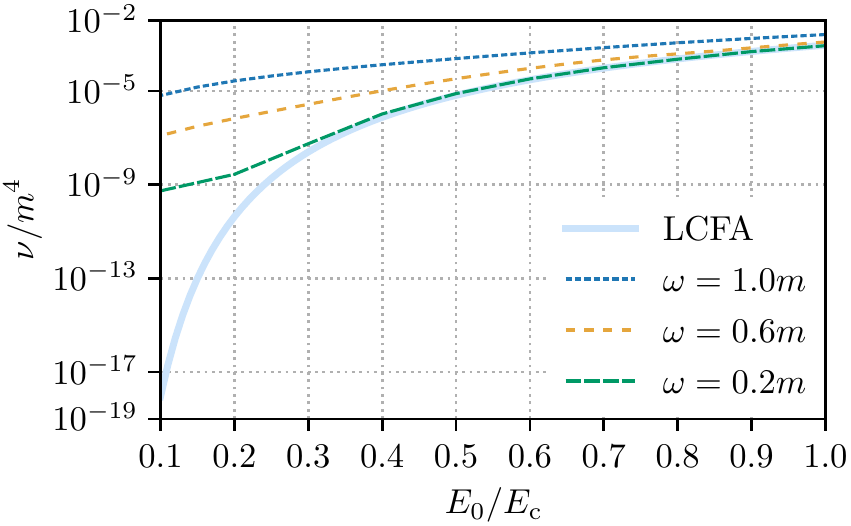}}
  \caption{Parameter $\nu = \omega N/V$ as a function of the amplitude of the standing wave~\eqref{eq:field_config} evaluated within the LCFA (solid light-blue line) and computed exactly for various values of the field frequency $\omega$ (dashed lines).}
  \label{fig:sw}
\end{figure}
%

\section{Standing electromagnetic wave} \label{sec:sw}

Here as in Sec.~\ref{sec:oef_num} we will compare the approximate $\omega$-independent results with those obtained by means of the approach described in Sec.~\ref{sec:furry} in the case of a standing-wave background~\eqref{eq:field_config}. This specific scenario properly takes into account the magnetic component of the external field according to Maxwell's equations, so it represents more realistic setup which mimics a combination of two counterpropagating laser pulses. Besides, as the external background is a periodic function of $z$, the computations can be carried out very efficiently, cf.~Ref.~\cite{aleksandrov_prd_2017_2}.

The results in terms of the function $\nu = \omega N/V$ are displayed in Fig.~\ref{fig:sw}. The overall behavior of the curves is very similar to that shown in Fig.~\ref{fig:oef}. First, the LCFA is accurate for sufficiently low field frequencies and large field amplitudes, while at small $E_0$ the discrepancy is tremendous. Second, although it is not clear from the graph, the LCFA begins to overestimate the particle yield for sufficiently large values of the field amplitude. The onset of this overestimation is shifted compared to the case of an oscillating electric field since the spatial cosine-profile of a standing wave makes the field effectively weaker, so the Pauli principle plays a significant role for larger values of the parameter $E_0$. In other words, in the case of a homogeneous background, the root mean square value of the field strength coincides with $E_0$, while in the case of a standing wave, it is $\sqrt{2}$ times smaller than $E_0$. We find that for these two scenarios, the dashed curve for $\omega = 0.2m$ intersects the LCFA one at different values of $E_0$, whose ratio approximately matches $\sqrt{2}$. Third, one can directly verify that according to our findings, the validity of the LCFA is not governed by the condition $\gamma \ll 1$.

The similarity between Figs.~\ref{fig:oef} and \ref{fig:sw} indicates that the applicability of the LCFA is not sensitive to the specific form of the field profile, which is quite natural as we only count the number of particles without following their dynamics in the external field (the latter is extremely important for predicting the momentum spectra~\cite{aleksandrov_prd_2019_1, aleksandrov_kohlfuerst}).

Let us finally provide a simple example of how one can assess the applicability of the LCFA. To this end, we consider a field configuration involving multiple laser pulses~\cite{bulanov_prl_2010}. The characteristic field amplitude chosen in Ref.~\cite{bulanov_prl_2010} is $E_0 \simeq (0.1\text{--}0.2)E_\text{c}$, while the field frequency is very small, $\omega \sim 10^{-6} \, m$. The direct comparison between the LCFA and exact calculations displayed in Figs.~\ref{fig:oef} and \ref{fig:sw} clearly indicates that for such low frequencies the LCFA should be justified. On the other hand, at $E_0 \sim 0.1 E_\text{c}$, one can also employ the requirement~\eqref{eq:LCFA_SC_cond_2}, which in our case demands $10^4 \gg 1$. Since this condition is also satisfied, the LCFA analysis performed in Ref.~\cite{bulanov_prl_2010} is evidently reliable.

\section{Conclusion}\label{sec:conclusion}

In the present investigation, we scrutinized the validity of the locally constant field approximation by comparing its predictions with the exact values of the total number of $e^+ e^-$ pairs. In the region of large field amplitudes ($E_0 \gtrsim 0.1 E_\text{c}$) and high frequencies ($\omega \gtrsim 0.1m$), we performed numerical computations by means of two nonperturbative techniques. It was shown that the LCFA may indeed be very accurate even if the Keldysh parameter is close to $\gamma = 1/2$, i.e., the condition $\gamma \ll 1$ is not satisfied. On the other hand, for $E_0 \lesssim 0.1 E_\text{c}$ and $\omega \lesssim 0.1m$, even the requirement $\gamma \ll 1$ does not necessarily justify the LCFA. In fact, one has to take care that the parameter $|eE_0|^{3/2}/(\omega m^2) = (E_0/E_\text{c})^{3/2} (m/\omega)$ is sufficiently large. This condition was deduced in the case of a Sauter pulse and that of an oscillating electric field, which suggests that the accuracy of the LCFA is governed by one universal parameter which differs from $\gamma$. 

Besides, we carried out exact calculations in the case of a standing electromagnetic wave depending both on time and one of the spatial coordinates. It was demonstrated that the main patterns revealed for an oscillating electric background hold also for a standing wave. This also corroborates that the LCFA justification is not that sensitive to the details of the spatio-temporal shape of the external field, although the particle dynamics could change drastically. In this study, we considered a standing wave whose ``temporal'' frequency $\omega$ matches the frequency regarding the coordinate dependence $k_z$ according to Maxwell's equations. One may expect that in a hypothetical scenario involving a standing wave with $k_z \neq \omega$, one should simply choose the higher frequency in order to find out whether the LCFA is applicable. This issue is, however, beyond the scope of the present study.

Our investigation provides a theoretical basis which allows one to employ the LCFA and perform very efficient computations as soon as the parameters of the external field configuration are chosen within the proper domain identified in this paper.

\begin{acknowledgments}
This work was supported by Russian Foundation for Basic Research (RFBR) (Grant No.~20-52-12017) and by Deutsche Forschungsgemeinschaft (DFG) (Grant No.~PL~254/10-1). D.G.S. acknowledges also the support from the St. Petersburg University Alumni Association.
\end{acknowledgments}

\appendix*

\section{LCFA expression in the case of an oscillating electric field}

Here we will represent the integral in the LCFA expression~\eqref{eq:LCFA_uniform_wo_envelope} in the form~\eqref{eq:LCFA_J_def} and derive the asymptotic expansion~\eqref{eq:LCFA_mu_asym}. First, we use the substitution $1/\cos \phi = y$ to obtain the following formula:
\begin{equation}
 \frac{N_{\infty}^{\text{(LCFA)}}}{VT} = \frac{m^4}{2\pi^4} \, \bigg ( \frac{E_0}{E_\text{c}} \bigg )^2 \mathcal{J} \bigg ( \frac{\pi E_\text{c}}{E_0} \bigg ),
 \label{eq:app_first}
\end{equation}
where
\begin{equation}
\mathcal{J} (z) = \int \limits_1^{\infty} \frac{\mathrm{e}^{-zy}}{y^3 \sqrt{y^2-1}} \, dy.
 \label{eq:app_J_def}
\end{equation}
This function can be represented as
\begin{equation}
\mathcal{J} (z) = \int \limits_1^{\infty} \frac{\mathrm{e}^{-zy}}{y \sqrt{y^2-1}} \, dy - \int \limits_1^{\infty} \frac{\sqrt{y^2-1}}{y^3} \, \mathrm{e}^{-zy} \, dy.
 \label{eq:app_J_I}
\end{equation}
Using the identity
\begin{equation}
\frac{\mathrm{e}^{-zy}}{y} = \int \limits_z^{\infty} \mathrm{e}^{-yx} \, dx,
 \label{eq:app_identity}
\end{equation}
we write the first term in Eq.~\eqref{eq:app_J_I} in the form
\begin{equation}
\mathcal{I}_1 (z) =  \int \limits_z^{\infty} K_0 (x) dx,
 \label{eq:app_I1}
\end{equation}
where $K_\nu (x)$ are the modified Bessel functions of the second kind. The second integral in Eq.~\eqref{eq:app_J_I} can be rewritten via integration by parts:
\begin{equation}
\mathcal{I}_2 (z) = -\frac{z}{2} \, \int \limits_1^{\infty} \frac{\sqrt{y^2-1}}{y^2} \, \mathrm{e}^{-zy} \, dy + \frac{1}{2} \, \mathcal{I}_1(z).
 \label{eq:app_I2}
\end{equation}
Integrating by parts one more time and using Eq.~\eqref{eq:app_identity}, we arrive at
\begin{equation}
\mathcal{I}_2 (z) = \frac{z^2}{2} \, \int \limits_z^{\infty} \frac{K_1(x)}{x} \, dx - \frac{z}{2} \, K_0 (z) + \frac{1}{2} \, \mathcal{I}_1(z).
 \label{eq:app_I2_K}
\end{equation}
Now using the identity $K_1 (x) = - K'_0 (x)$, we obtain
\begin{equation}
\mathcal{I}_2 (z) = -\frac{z^2}{2} \, \int \limits_z^{\infty} \frac{K_0(x)}{x^2} \, dx + \frac{1}{2} \, \mathcal{I}_1(z).
 \label{eq:app_I2_final}
\end{equation}
Combining Eqs.~\eqref{eq:app_I1} and \eqref{eq:app_I2_final}, we present $\mathcal{J} (z) = \mathcal{I}_1(z) - \mathcal{I}_2(z)$ in the form
\begin{equation}
\mathcal{J} (z) = \frac{1}{2} \, \int \limits_z^{\infty} \bigg ( 1 + \frac{z^2}{x^2} \bigg ) K_0 (x) dx
 \label{eq:app_J_final_1}
\end{equation}
or, equivalently,
\begin{equation}
\mathcal{J} (z) = \frac{z}{2} \, \int \limits_1^{\infty} \bigg ( 1 + \frac{1}{x^2} \bigg ) K_0 (zx) dx,
 \label{eq:app_J_final_2}
\end{equation}
which coincides with Eq.~\eqref{eq:LCFA_J_def}. In order to identify the asymptotic behavior~\eqref{eq:LCFA_mu_asym}, we assume that $z \gg 1$ according to Eq.~\eqref{eq:app_first} and employ the asymptotic expansion of the modified Bessel function $K_0$,
\begin{equation}
K_0 (x) = \sqrt{\frac{\pi}{2x}} \, \mathrm{e}^{-x} \bigg [ 1 - \frac{1}{8x} + \frac{9}{128x^2} + \mathcal{O} \bigg ( \frac{1}{x^3} \bigg ) \bigg ].
 \label{eq:app_K0_asym}
\end{equation}
Performing then integration in Eq.~\eqref{eq:app_J_final_2}, one receives an expression involving incomplete Gamma functions $\Gamma (a, z)$. By means of the recurrence relation
\begin{equation}
\Gamma(a, z) = e^{-z} z^{a-1} + (a-1)\, \Gamma (a-1, z)
 \label{eq:app_Gamma_rec}
\end{equation}
or using directly the formula
\begin{equation}
\Gamma (a, z) = z^{a-1} \, \mathrm{e}^{-z} \bigg [ 1 - \frac{1-a}{z} + \frac{(2-a)(1-a)}{z^2} + \mathcal{O} \bigg ( \frac{1}{z^3} \bigg ) \bigg ],
 \label{eq:app_Gamma_asym}
\end{equation}
we collect the terms and obtain
\begin{equation}
\mathcal{J} (z) = \sqrt{\frac{\pi}{2z}} \, \mathrm{e}^{-z} \bigg [ 1 - \frac{13}{8z} + \frac{657}{128z^2} + \mathcal{O} \bigg ( \frac{1}{z^3} \bigg ) \bigg ].
 \label{eq:app_J_asym}
\end{equation}
Taking into account Eqs.~\eqref{eq:app_first}, \eqref{eq:app_J_asym}, \eqref{eq:LCFA_uniform_mu_def}, and \eqref{eq:mu_SC}, one immediately arrives at Eq.~\eqref{eq:LCFA_mu_asym}.

\bibliography{list}

\end{document}